\documentclass[11pt,a4paper]{article}
\usepackage[utf8]{inputenc}
\usepackage[T1]{fontenc}
\usepackage{lmodern}
\usepackage[margin=2.3cm]{geometry}
\usepackage{amsmath,amssymb}
\usepackage{graphicx}
\usepackage{booktabs}
\usepackage{tabularx}
\usepackage{xcolor}
\usepackage{tikz}
\usetikzlibrary{arrows.meta,positioning,fit,calc}
\usepackage{enumitem}
\usepackage[hidelinks]{hyperref}
\usepackage{url}
 
\setlist{itemsep=1pt,topsep=3pt}
 
\newcommand{\repo}{\url{https://github.com/ZeboZebo702/tt_matmul_2x2_int8}}
 
\tikzset{
  blk/.style={draw,rounded corners=2pt,minimum height=7mm,minimum width=18mm,align=center,font=\small},
  reg/.style={draw,fill=gray!15,minimum height=6mm,minimum width=12mm,font=\small},
  new/.style={blk,fill=orange!15},
  arr/.style={-{Stealth[length=2mm]},thick},
  lbl/.style={font=\footnotesize,inner sep=1pt}
}
 
\title{MAC units for a 2$\times$2 int8 systolic array on Tiny Tapeout}
\author{Danila Gorodecky (Gorodetsky)\\ INESC-ID, Lisbon, Portugal\\ \texttt{danila.gorodecky@gmail.com}}
\date{October 2026}
 
\begin{document}
\maketitle
 
\begin{abstract}
We take the multiply-accumulate (MAC) unit of an open-source 2$\times$2 int8 systolic array taped out on Tiny Tapeout (GF180 process) and propose two replacements. The first one, \emph{hybrid}, uses a different arithmetic structure for each row of the array. It reduces the logic area of one MAC unit by 4--6\% and of the whole chip by 0.7--3.2\% after place and route. The second one, \emph{fused\_eq}, never builds the long carry chain of the accumulation. It shortens the critical path by 20\% for one MAC unit and by 14\% for the whole chip. All numbers come from the standard Tiny Tapeout flow (LibreLane) run in GitHub Actions. Correctness is checked by exhaustive simulation of all $2^{24}$ inputs, a formal equivalence proof, and the original test bench.
\end{abstract}
 
\section{Introduction}
 
Tiny Tapeout~\cite{tinytapeout} lets many people put small designs on one shared chip. Each design gets one or more \emph{tiles}. A tile is a fixed piece of silicon. On the GF180 process~\cite{gf180pdk} one tile is about $347\times161~\mu$m. Area is the main cost, so compact arithmetic blocks are useful for this community.
 
As a test case we use \texttt{tt\_um\_essen}, a 2$\times$2 systolic array for int8 matrix multiplication~\cite{essen_repo}. It was taped out on a Tiny Tapeout GF180 shuttle. A \emph{systolic array}~\cite{kung} is a grid of identical units. Data move from unit to unit at each clock cycle, so each value is reused many times. Each unit of this array is a  multiply–accumulate (MAC) unit: it multiplies two numbers and adds the result to a value from its neighbour.
 
We optimize the MAC unit for area or for speed. The function and the design stay the same as it is in the standard Tiny Tapeout flow with its default tools and settings. Only the arithmetic inside the unit changes. We describe two proposed architectures and compare them with the original on the real flow. We share the Verilog files of both units and all files needed to rerun the flow.
 
\section{The original design}
 
\subsection{The array}
 
\begin{figure}[t]
\centering
\begin{tikzpicture}[node distance=12mm and 16mm]
\node[blk,minimum width=20mm,minimum height=11mm] (m00) {MAC (0,0)\\ $w_{00}$};
\node[blk,minimum width=20mm,minimum height=11mm,right=of m00] (m10) {MAC (1,0)\\ $w_{10}$};
\node[blk,minimum width=20mm,minimum height=11mm,below=of m00] (m01) {MAC (0,1)\\ $w_{01}$};
\node[blk,minimum width=20mm,minimum height=11mm,below=of m10] (m11) {MAC (1,1)\\ $w_{11}$};
\draw[arr] ($(m00.west)+(-12mm,0)$) node[left,lbl]{$i_0$} -- (m00);
\draw[arr] ($(m01.west)+(-12mm,0)$) node[left,lbl]{$i_1$} -- (m01);
\draw[arr] (m00) -- (m10); \draw[arr] (m01) -- (m11);
\draw[arr] ($(m00.north)+(0,9mm)$) node[above,lbl]{0} -- (m00);
\draw[arr] ($(m10.north)+(0,9mm)$) node[above,lbl]{0} -- (m10);
\draw[arr] (m00) -- node[right,lbl]{8 bit} (m01); \draw[arr] (m10) -- node[right,lbl]{8 bit} (m11);
\draw[arr] (m01.south) -- ++(0,-8mm) node[below,lbl]{$r_0$};
\draw[arr] (m11.south) -- ++(0,-8mm) node[below,lbl]{$r_1$};
\node[lbl,anchor=west,align=left] at ($(m10.east)+(8mm,0)$) {\textbf{top row}\\ $c=0$ always};
\node[lbl,anchor=west,align=left] at ($(m11.east)+(8mm,0)$) {\textbf{bottom row}\\ $c$ = result of the unit above};
\end{tikzpicture}
\caption{The 2$\times$2 systolic array of \texttt{tt\_um\_essen}. The top row always gets 0 from above.}
\label{fig:array}
\end{figure}
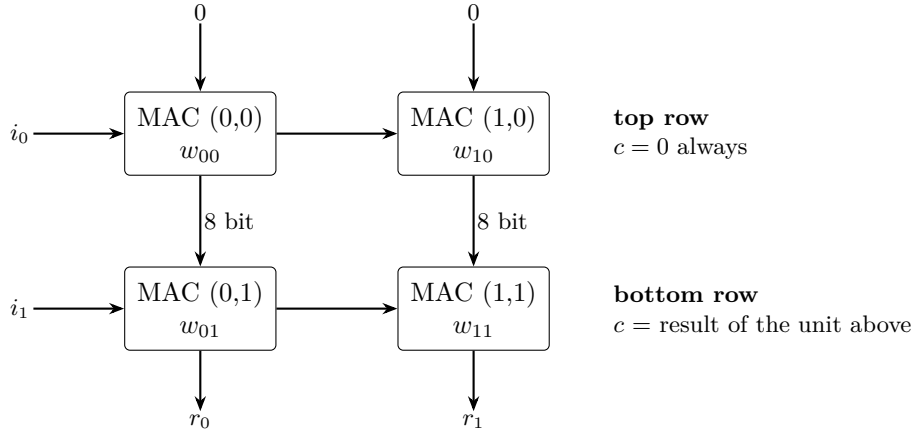

Fig.~\ref{fig:array} represents the array, where input values $i$ move to the right, weights $w$ are stored inside the units, partial sums move down, the units in the top row get 0 from above, the bottom row gives the results.
 
All numbers in the array are \emph{int8}: signed integers of 8 bits. They are stored in \emph{two's complement}, the usual way to store negative numbers in hardware. The rule is simple. Read the 8 bits as an ordinary binary number from 0 to 255. If the top bit (bit 7) is 0, this is the value. If the top bit is 1, subtract 256. So the top bit is the sign, and the range is from $-128$ to $127$. Table~\ref{tab:int8} gives examples.
 
\begin{table}[h]
\centering\small
\begin{tabular}{ccc}
\toprule
8 bits & as an unsigned number & int8 value \\
\midrule
\texttt{00000101} & 5 & 5 \\
\texttt{01111111} & 127 & 127 (the largest) \\
\texttt{10000000} & 128 & $128-256=-128$ (the smallest) \\
\texttt{11111101} & 253 & $253-256=-3$ \\
\texttt{11111111} & 255 & $255-256=-1$ \\
\bottomrule
\end{tabular}
\caption{How 8 bits are read as an int8 number.}
\label{tab:int8}
\end{table}
 
Each unit computes
\begin{equation}
r = \mathrm{clamp}(a\cdot w + c),\qquad \mathrm{clamp}(x)=\min(127,\max(-128,x)),
\label{eq:mac}
\end{equation}
where $a$ is the input, $w$ the weight, $c$ the value from above and $r$ the result. All four numbers $a$, $w$, $c$, $r$ are int8. The function $\mathrm{clamp}$ is \emph{saturation}: a value that does not fit in int8 becomes the nearest limit. There are three examples:
\begin{itemize}
\item $a=10$, $w=10$, $c=20$: $100+20=120$. It fits, so $r=120$.
\item $a=10$, $w=10$, $c=50$: $100+50=150>127$, so $r=127$.
\item $a=-100$, $w=2$, $c=0$: $-200<-128$, so $r=-128$.
\end{itemize}
The product $a\cdot w$ needs 16 bits, so without saturation the data width would grow at each step.
 
\subsection{The MAC unit}
 
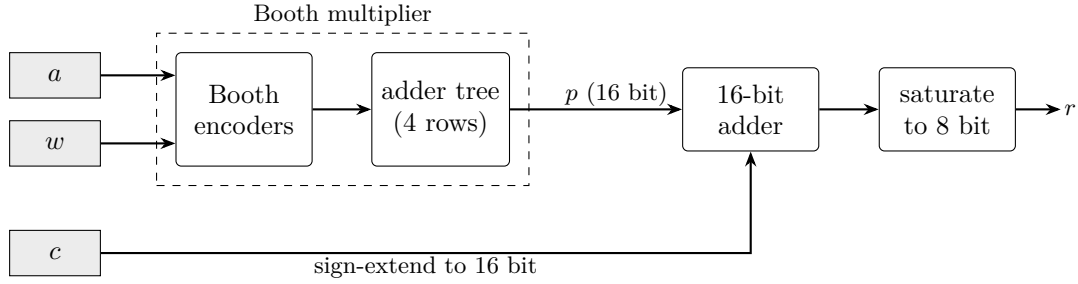
\begin{figure}[t]
\centering
\begin{tikzpicture}
\node[reg] (a) at (0,0.45) {$a$};
\node[reg] (w) at (0,-0.45) {$w$};
\node[reg] (c) at (0,-1.9) {$c$};
\node[blk,minimum height=15mm] (enc) at (2.5,0) {Booth\\ encoders};
\node[blk,minimum height=15mm] (tree) at (5.1,0) {adder tree\\ (4 rows)};
\node[blk,minimum height=11mm] (add) at (9.2,0) {16-bit\\ adder};
\node[blk,minimum height=11mm] (sat) at (11.8,0) {saturate\\ to 8 bit};
\draw[arr] (a) -- (a-|enc.west); \draw[arr] (w) -- (w-|enc.west);
\draw[arr] (enc) -- (tree);
\draw[arr] (tree) -- node[above,lbl,pos=0.62]{$p$ (16 bit)} (add);
\draw[arr] (c) -- node[below,lbl]{sign-extend to 16 bit} (c-|add) -- (add);
\draw[arr] (add) -- (sat); \draw[arr] (sat) -- ++(1.5,0) node[right,lbl]{$r$};
\node[draw,dashed,inner sep=2.5mm,fit=(enc)(tree)] (mul) {};
\node[lbl,above=0.5mm of mul] {Booth multiplier};
\end{tikzpicture}
\caption{The original MAC unit. The product and the accumulation use two separate carry-propagate adders in series.}
\label{fig:orig}
\end{figure}

Fig.~\ref{fig:orig} shows the original unit with three registers which hold $a$, $w$ and $c$. The product is computed by a radix-4 Booth multiplier~\cite{booth,parhami}. Booth encoding looks at the bits of $w$ in groups of three. Each group selects $0$, $\pm a$ or $\pm 2a$, so an 8-bit multiplier needs only four \emph{partial products} (shifted rows to be added) instead of eight. Hence an adder tree sums the four rows into a 16-bit product $p$.
 
Then a 16-bit adder computes $p+c$, with $c$ sign-extended to 16 bits. Finally, the saturation logic checks the upper bits of the sum. If bits 15 to 7 are all equal, the sum fits in 8 bits. Otherwise the result is 127 or $-128$. The whole unit works in one clock cycle. The chip was signed off at 50~MHz.
 
\section{Proposed units}
 
\subsection{9 bits of the product are enough}
 
The value $c$ is between $-128$ and $127$. If $p\ge 256$, then $p+c\ge 128$ and the result is 127 for any $c$. In the same way, if $p\le -257$, the result is $-128$ for any $c$. So we can replace $p$ by
\begin{equation}
p_9=\min(255,\max(-256,p)),
\label{eq:p9}
\end{equation}
which fits in 9 bits, and still get exactly $\mathrm{clamp}(p_9+c)=\mathrm{clamp}(p+c)$. We checked this for all $2^{24}$ triples $(a,w,c)$. A flag \emph{big} tells that $p$ does not fit in 9 bits (bits 15 to 8 of $p$ are not all equal). Then the result is chosen by the sign of $p$. In parallel, a short 10-bit adder computes $p_{[8:0]}+c$. We call this structure \emph{presat}.
 
\subsection{\textit{hybrid}: a different structure for each row}
 
The two units that get 0 from above form the \emph{top row}: MAC (0,0) and MAC (1,0) in Fig.~\ref{fig:array}. For them $c=0$ always. All other units get $c$ from the unit above, so $c$ can be any int8 value. In the 2$\times$2 array these are the two units of the bottom row, in a larger array they would be all rows except the first.
 
Synthesis tools recognize the constant $c=0$ and remove the parts of the circuit that add $c$. So a top-row unit is much smaller: 9\,562~$\mu$m$^2$ against 12\,697~$\mu$m$^2$ for the original (Table~\ref{tab:explore}). For the same reason the structure that is best for one row is not the best for the other:
\begin{itemize}
\item top row ($c=0$): \emph{presat} gives 9\,349~$\mu$m$^2$, 2.2\% less than the original. \emph{tree} (described below) gives 9\,687~$\mu$m$^2$, 1.3\% more;
\item other rows (any $c$): \emph{tree} gives 12\,078~$\mu$m$^2$, 4.9\% less. \emph{presat} gives 12\,539~$\mu$m$^2$, only 1.2\% less.
\end{itemize}
So \emph{hybrid} uses \textit{presat} for the top row and \textit{tree} for the other rows.
 
\begin{table}[t]
\centering\small
\begin{tabular}{lrr}
\toprule
Structure & top row, $c=0$ & other rows, any $c$ \\
\midrule
original & 9\,562 & 12\,697 \\
presat & 9\,349 ($-2.2\%$) & 12\,539 ($-1.2\%$) \\
tree & 9\,687 ($+1.3\%$) & 12\,078 ($-4.9\%$) \\
\bottomrule
\end{tabular}
\caption{Synthesis area of one MAC unit in $\mu$m$^2$. No structure is the best in both columns, which is why hybrid uses two of them.}
\label{tab:explore}
\end{table}

In \emph{tree} (Fig.~\ref{fig:hybrid}, bottom) the four Booth rows and $c$ are written as one sum with several terms. Yosys~\cite{yosys} then builds one adder tree for the whole $a\cdot w+c$, without a separate adder for $c$. To add an 8-bit signed $c$ to a 16-bit sum, $c$ must first be \emph{sign-extended}: its top bit is copied into bits 8 to 15. For example, $-3=\mathtt{11111101}$ becomes \texttt{1111111111111101}. These eight copies are extra inputs of the adder tree. A standard trick removes them:
\begin{equation}
c = c' - 128,\qquad c' = c \oplus \mathtt{0x80}.
\label{eq:trick}
\end{equation}
Here \texttt{0x80} is the number 80 written in hexadecimal (base 16), that is $8\cdot16=128$, or \texttt{10000000} in binary. The sign $\oplus$ is bitwise XOR: a bit XOR 1 is inverted, a bit XOR 0 stays the same. So $c'$ is $c$ with only the top bit inverted. We read $c'$ as an unsigned number from 0 to 255. Two examples:
\begin{itemize}
\item $c=-3=\mathtt{11111101}$. Then $c'=\mathtt{01111101}=125$, and $125-128=-3$.
\item $c=5=\mathtt{00000101}$. Then $c'=\mathtt{10000101}=133$, and $133-128=5$.
\end{itemize}
$c'$ is unsigned, so its bits 8 to 15 are 0 and it takes only 8 bits in the tree. The term $-128$ is a constant. A constant costs almost nothing, because the tools simplify every adder input that is always 0 or always 1. So the tree gets an 8-bit row instead of a 16-bit row.
 
The \emph{hybrid} unit has a parameter \texttt{TOP\_ROW}. If it is 1, the unit uses \emph{presat}; otherwise it uses \emph{tree}. Both structures are correct for any input, so each setting is bit-exact on its own. The only change outside the unit is one line in the array file \texttt{mac.v}: \texttt{mac\_unit \#(.W(W), .TOP\_ROW(y == 0))}.
 
\begin{figure}[t]
\centering
\begin{tikzpicture}
% top row
\node[font=\small\bfseries,anchor=west] at (0,1.6) {Top row, $c=0$ (\texttt{TOP\_ROW}=1): presat};
\node[blk,minimum height=11mm] (m1) at (1.3,0) {Booth\\ multiplier};
\node[new,minimum width=30mm] (big) at (5.2,0.75) {flag \emph{big}:\\ $p$ does not fit in 9 bit};
\node[new,minimum width=30mm] (a10) at (5.2,-0.75) {10-bit adder\\ $p_{[8:0]}+c$};
\node[blk,minimum height=11mm] (mx1) at (10.0,0) {select and\\ saturate};
\draw[arr] (m1.east) -- ++(0.5,0) |- (big.west);
\draw[arr] (m1.east) -- ++(0.5,0) |- (a10.west);
\draw[arr] (big.east) -- ++(0.6,0) |- ([yshift=2mm]mx1.west);
\draw[arr] (a10.east) -- ++(0.6,0) |- ([yshift=-2mm]mx1.west);
\draw[arr] (mx1) -- ++(1.6,0) node[right,lbl]{$r$};
% other rows
\node[font=\small\bfseries,anchor=west] at (0,-2.0) {Other rows, any $c$ (\texttt{TOP\_ROW}=0): tree};
\node[blk,minimum height=14mm] (e2) at (1.3,-3.5) {Booth\\ encoders};
\node[new,minimum width=36mm,minimum height=14mm] (tr2) at (5.2,-3.5) {one adder tree:\\ 4 Booth rows $+\;c'$\\ + constant $-128$};
\node[blk,minimum height=11mm] (s2) at (10.0,-3.5) {saturate\\ to 8 bit};
\draw[arr] (e2) -- (tr2);
\draw[arr] (tr2) -- node[above,lbl]{$a\cdot w+c$} (s2);
\draw[arr] (s2) -- ++(1.6,0) node[right,lbl]{$r$};
\end{tikzpicture}
\caption{The hybrid MAC unit. Orange blocks are new. $c'$ is $c$ with the top bit inverted, Eq.~(\ref{eq:trick}). Both structures compute Eq.~(\ref{eq:mac}) for any input.}
\label{fig:hybrid}
\end{figure}
 
\subsection{\textit{fused\_eq}: no long carry chain}
 
The critical path of the original unit goes through two carry-propagate adders in series: the last adder of the multiplier and the 16-bit accumulation adder. A carry may have to propagate through all 16 bits of each of them.
 
\begin{figure}[t]
\centering
\begin{tikzpicture}
\node[blk,minimum height=18mm] (e) at (1.0,0) {Booth\\ encoders};
\node[new,minimum width=32mm,minimum height=18mm] (csa) at (4.3,0) {3:2 compressor tree\\ 6 rows:\\ 4 rows $+\,c$ + constant};
\node[new,minimum width=30mm] (lo) at (9.0,1.25) {8-bit adder (low bits)};
\node[new,minimum width=30mm] (fit) at (9.0,0) {fits in int8? (bitwise)};
\node[new,minimum width=30mm] (sg) at (9.0,-1.25) {sign of $a\cdot w$};
\node[blk,minimum height=33mm] (mx) at (12.4,0) {select};
\draw[arr] (e) -- (csa);
\draw[arr] (csa.east) -- ++(0.8,0) coordinate (j) |- (lo.west);
\draw[arr] (j) -- (fit.west);
\node[lbl,below] at ($(csa.east)!0.5!(j)$) {$X,Y$};
\draw[arr] (lo.east) -- (lo.east-|mx.west);
\draw[arr] (fit) -- (mx);
\draw[arr] (sg.east) -- (sg.east-|mx.west);
\draw[arr] ($(sg.south)+(0,-0.55)$) node[below,lbl]{signs of $a$ and $w$} -- (sg.south);
\draw[arr] (mx) -- ++(1.3,0) node[right,lbl]{$r$};
\end{tikzpicture}
\caption{The \textit{fused\_eq} MAC unit. The three orange blocks on the right work in parallel.}
\label{fig:fused}
\end{figure}
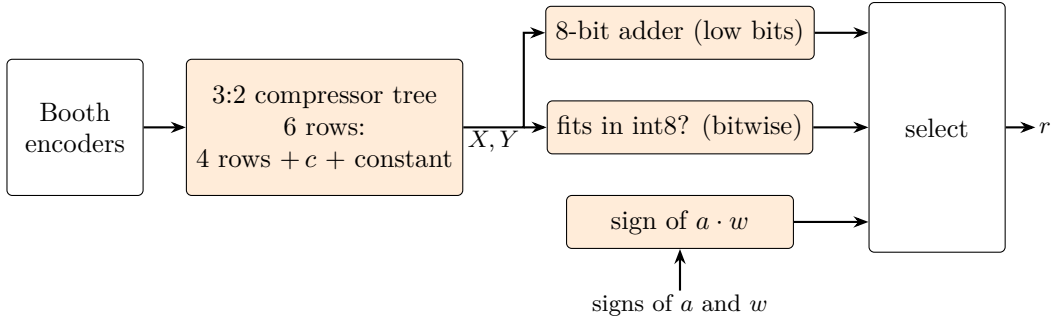

In \emph{fused\_eq} (Fig.~\ref{fig:fused}) $c$ is added as one more row of the multiplier. Three levels of 3:2 compressors (full adders that turn three rows into two without carry propagation~\cite{wallace}) reduce the six rows to two numbers $X$ and $Y$ with $X+Y=a\cdot w+c$ modulo $2^{16}$. The long sum $X+Y$ is never computed, instead of this:
\begin{itemize}
\item an 8-bit adder gives the low 8 bits of the result and the carry into bit 7;
\item a check responses if bits 15 to 7 of $X+Y$ are all 0 or all 1, that is, if the sum fits in int8. For bit $k$ of a sum $S=X+Y$ we have $S_k = X_k\oplus Y_k\oplus C_k$, where $C_k$ is the carry into bit $k$. If all bits of $S$ must be 0, then the needed carry is $C_{k+1}=X_k\lor Y_k$; if all bits must be 1, then $C_{k+1}=X_k\land Y_k$. So each bit can be checked on its own, in parallel, with no carry chain;
\item the direction of overflow is the sign of the product. It is known early from the signs of $a$ and $w$ and from a zero check.
\end{itemize}
A multiplexer then selects the 8-bit sum, 127 or $-128$.
 
\subsection{The difference}
\label{sec:jtag}
 
The original unit has a debug register that can be read through JTAG. Address 3 returns the upper byte of the 16-bit sum. The new units never build this sum, so address 3 returns the saturation flags instead. Addresses 0 to 2 and all computed results are the same.
 
\section{Measurements}
 
\subsection{Flow and settings}
 
We use the flow Tiny Tapeout uses (Table~\ref{tab:versions}): the tt-support-tools scripts~\cite{tt_tools} and LibreLane~\cite{librelane}. LibreLane runs synthesis with Yosys and ABC~\cite{yosys,abc}, then floorplan, placement, clock tree, routing and static timing analysis (STA) with OpenROAD~\cite{openroad}. STA computes the delay of every path with the real wires. All settings are the default Tiny Tapeout settings for GF180 at 3.3~V. Only the clock period changes between runs. All runs are done in GitHub Actions, a GitHub service that runs scripts on its own virtual machines.
 
\begin{table}[t]
\centering\small
\begin{tabular}{ll}
\toprule
Item & Version \\
\midrule
Original design & \texttt{Essenceia/Systolic\_MAC\_with\_DFT}, commit \texttt{6a3c04e} \\
Tiny Tapeout tools & \texttt{tt-support-tools}, commit \texttt{01d5d28} \\
Flow & LibreLane 3.0.14 (in its Docker image) \\
Process & GF180MCU D, cell library \texttt{gf180mcu\_fd\_sc\_mcu7t5v0}, 3.3 V \\
Computers & GitHub Actions virtual machines, Ubuntu 24.04 \\
\bottomrule
\end{tabular}
\caption{Fixed versions. All variants, including the original, use the same versions.}
\label{tab:versions}
\end{table}
 
\subsection{Two levels}
 
\begin{itemize}
\item \textbf{One unit.} A small wrapper puts one MAC unit on a 1$\times$1 tile. The timed path is the same as in the chip: input registers, multiplier, adder, saturation, output register. Little else is on the tile, so the difference between variants is the difference of the arithmetic.
\item \textbf{Whole chip.} The full \texttt{tt\_um\_essen} project on its 2 tiles, with only \texttt{mac\_unit.v} replaced (and one line of \texttt{mac.v} for hybrid). This shows the real effect on the project.
\end{itemize}
 
Each level runs at clock periods of 40, 30 and 20~ns (20~ns is 50~MHz).
 
\subsection{Metrics}
 
\begin{itemize}
\item \textbf{Logic area}: the sum of the areas of all cells after routing (gates, flip-flops, buffers), without fill, tap, end-cap, decap and antenna cells. These cells only fill empty space and depend on the tile, not on the design.
\item \textbf{Synthesis area}: the cell area right after synthesis, before the tools add buffers and resize cells for timing.
\item \textbf{Critical path}: clock period minus the worst setup slack. \emph{Slack} is the time left before the end of the clock cycle. We report the typical corner (25~$^\circ$C, 3.3~V). A \emph{corner} is a set of process, voltage and temperature conditions. Tiny Tapeout signs off designs at this corner.
\end{itemize}
 
\subsection{Correctness}
\label{sec:checks}
 
Every variant passed all three checks below before any area or timing run. The checks test different things.
\begin{enumerate}
\item \textbf{Exhaustive simulation} with Icarus Verilog~\cite{iverilog}. It tests the arithmetic of one MAC unit. All $2^{24}$ combinations of $(a,w,c)$ are applied. The original and all variants run side by side and must give the same $r$. The original is also compared with Eq.~(\ref{eq:mac}) computed directly.
\item \textbf{Formal equivalence} with Yosys (\texttt{equiv\_make}, \texttt{equiv\_simple}, \texttt{equiv\_induct}). This is a proof, not a test: it uses no input values. The registers have the same names in both designs. The tool proves that the logic between them gives the same outputs for every possible register value. JTAG address 3 is excluded (Section~\ref{sec:jtag}).
\item \textbf{The original test bench}: the 7 cocotb~\cite{cocotb} tests of the original project. They test the unit inside the whole chip: the control logic, the data flow between the four units and JTAG. For hybrid they also test the changed \texttt{mac.v}.
\end{enumerate}
Checks 1 and 2 overlap on purpose. Two different tools give the same answer, so an error in one tool is unlikely to hide a bug. For \textit{hybrid}, both are done with both values of \texttt{TOP\_ROW}.
 
\subsection{Repeatability}
 
The flow is deterministic and two separate sweeps gave identical results. On the other hand, the result of synthesis is sensitive to the structure of the input. Swapping two \texttt{always} blocks in the original code, which does not change the circuit, moved the synthesis area of the whole chip by 0.6\%. So we do not treat differences below about 1\% on the whole chip as real.
 
\section{Results}
 
\begin{figure}[t]
\centering
\includegraphics[width=\textwidth]{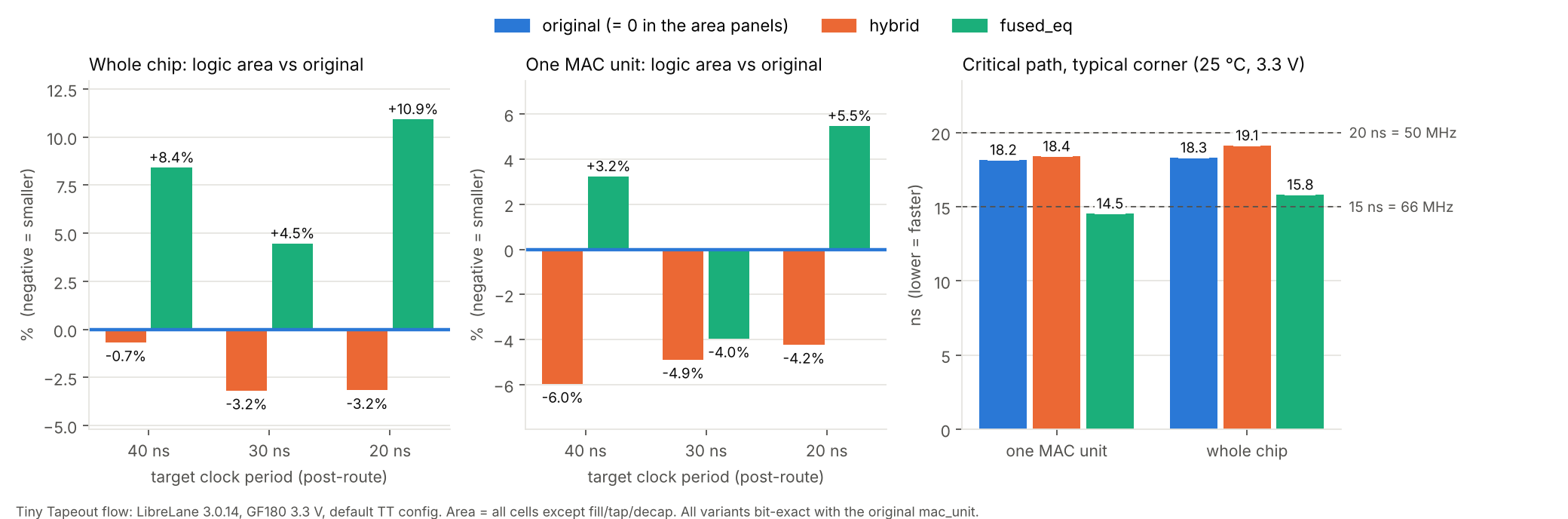}
\caption{Area change against the original (left, middle) and critical path (right).}
\label{fig:chart}
\end{figure}

Table~\ref{tab:area} gives the logic area and Table~\ref{tab:timing} the critical path. Fig.~\ref{fig:chart} shows both.
 
\begin{table}[t]
\centering\small
\begin{tabular}{llrrrr}
\toprule
Level & Variant & 40 ns & 30 ns & 20 ns & synthesis \\
\midrule
One unit & original & 16\,170 & 17\,221 & 17\,105 & 12\,697 \\
 & hybrid & 15\,204 ($-6.0\%$) & 16\,378 ($-4.9\%$) & 16\,381 ($-4.2\%$) & 12\,078 ($-4.9\%$) \\
 & fused\_eq & 16\,695 ($+3.2\%$) & 16\,539 ($-4.0\%$) & 18\,042 ($+5.5\%$) & 13\,351 ($+5.2\%$) \\
\midrule
Whole chip & original & 68\,124 & 71\,210 & 71\,089 & 54\,022 \\
 & hybrid & 67\,656 ($-0.7\%$) & 68\,940 ($-3.2\%$) & 68\,839 ($-3.2\%$) & 53\,330 ($-1.3\%$) \\
 & fused\_eq & 73\,860 ($+8.4\%$) & 74\,400 ($+4.5\%$) & 78\,863 ($+10.9\%$) & 57\,927 ($+7.2\%$) \\
\bottomrule
\end{tabular}
\caption{Logic area in $\mu$m$^2$ after place and route at three clock periods, and after synthesis. On the one-unit level hybrid is the same as tree.}
\label{tab:area}
\end{table}
 
\begin{table}[t]
\centering\small
\begin{tabular}{lrrr}
\toprule
Level & original & hybrid & fused\_eq \\
\midrule
One unit & 18.2 ns & 18.4 ns & 14.5 ns ($-20\%$) \\
Whole chip & 18.3 ns & 19.1 ns & 15.8 ns ($-14\%$) \\
\bottomrule
\end{tabular}
\caption{Critical path at the typical corner, 20 ns run. All variants meet 50 MHz. Only fused\_eq meets 15 ns (66 MHz) on the one-unit level; the original reaches 55 MHz.}
\label{tab:timing}
\end{table}
 
In short:
\begin{itemize}
\item \textit{hybrid} is the smaller unit: $-4$ to $-6\%$ of logic area for one unit, $-0.7$ to $-3.2\%$ for the whole chip. Its critical path is about the same as the original;
\item \textit{fused\_eq} is the faster unit: the critical path is 20\% shorter for one unit and 14\% shorter for the whole chip. It costs 4.5\% to 11\% more area for the whole chip; for one unit the change is from $-4\%$ to $+5.5\%$.
\end{itemize}
 
\section{Why the area gain is small}
 
\begin{enumerate}
\item \textbf{The baseline is good.} The original uses a hand-written radix-4 Booth multiplier, and for 8 bits Yosys and ABC are already close to the optimum.
\item \textbf{Arithmetic is less than half of the chip.} We replaced the arithmetic of the four MAC units with a stub that keeps all registers. The synthesis area of the chip dropped from 54\,k to 30\,k$\mu$m$^2$. So the arithmetic is about 45\% of the chip. The rest is flip-flops, the JTAG and design-for-test logic, and control.
\item \textbf{Half of the units are already simplified.} The top row gets a constant 0, and the tools remove most of the addition there by themselves.
\item \textbf{The tile count does not depend on the arithmetic.} One tile is about 55.7\,k$\mu$m$^2$, but cells can fill only about 60--70\% of it; the rest is for wires and power. The flow reports 67\% fill for the original chip on 2 tiles, so one tile holds about 34\,k$\mu$m$^2$ of cells. The chip has about 68\,k$\mu$m$^2$. Even with no MAC arithmetic at all it would have about 37\,k$\mu$m$^2$, still more than one tile.
\end{enumerate}
 
For larger arrays and wider number formats the arithmetic becomes the main part of the chip, the top row is a smaller share of the units, and the gain of one unit turns into the gain of the chip.

\paragraph{Slow corner.} At the slow corner (125~$^\circ$C, 3.0~V) no variant, including the original, meets its clock period in any run. The critical path there is 29--46~ns: the shorter the clock period, the harder the tools work and the shorter this path becomes. It does not affect the comparison, since all variants are measured the same way. It means that the place-and-route tools always try to repair this corner, and this makes the post-route area vary from one clock period to another. The synthesis area is free of this effect.
 
\section{Reproducing and reusing the setup}
 
All files are at \repo. Table~\ref{tab:repo} lists the main folders and files.

To repeat the study:
\begin{enumerate}
\item fork the repository, that is, make your own copy of it on GitHub;
\item in the Actions tab, start the \texttt{sweep} workflow. Choose the levels, the variants and the clock periods;
\item GitHub runs one job for each point (level, variant, clock period). The job downloads the original design at the fixed commit, puts in the variant, sets the clock period and runs the Tiny Tapeout hardening, the full flow from Verilog to layout;
\item a last job collects all results into tables, a CSV file and charts.
\end{enumerate}

\begin{table}[h]
\centering\small
\begin{tabularx}{\textwidth}{lX}
\toprule
Folder or file & What it does \\
\midrule
\texttt{variants/} & One file \texttt{mac\_unit\_<name>.v} per variant. \\
\texttt{variants/extra/} & Other changed files of a variant (for hybrid: \texttt{mac.v}). \\
\texttt{bench\_single/} & A small wrapper that puts one MAC unit on a 1$\times$1 tile. \\
\texttt{verify/} & Test benches and the formal proof script (Section~\ref{sec:checks}). \\
\texttt{scripts/verify.sh} & Runs all correctness checks. \\
\texttt{scripts/prepare.sh} & Builds one project: the original design + one variant + one clock period. \\
\texttt{scripts/run\_one.sh} & Runs \texttt{prepare.sh}, then the LibreLane flow, then collects the numbers of this run. \\
\texttt{scripts/report.py} & Makes the tables, a CSV file and the charts from all runs. \\
\texttt{scripts/synth\_local.py} & Runs only the synthesis step of LibreLane, without Docker. \\
\texttt{.github/workflows/sweep.yaml} & The experiment on GitHub: all chosen levels, variants and clock periods. \\
\texttt{.github/workflows/checks.yaml} & The correctness checks on GitHub. \\
\texttt{versions.env} & The fixed versions of Table~\ref{tab:versions}. \\
\bottomrule
\end{tabularx}
\caption{Main folders and files of the repository.}
\label{tab:repo}
\end{table}
 
To test a new block, put a new \texttt{mac\_unit\_<name>.v} in \texttt{variants/} and add its name to the list of variants. For any other Tiny Tapeout project, change the repository and commit in \texttt{versions.env} and the file that is replaced in \texttt{prepare.sh}.
 
For quick experiments, \texttt{synth\_local.py} runs only the synthesis step of LibreLane, with the same script and settings, in a few seconds. Its synthesis area matches the full flow within 0.1~$\mu$m$^2$ on the one-unit level.
 
\section{Conclusion}
 
Two drop-in MAC units for the \texttt{tt\_um\_essen} systolic array give the same results as the original. \emph{hybrid} makes one unit 4--6\% smaller and the whole chip up to 3.2\% smaller. \emph{fused\_eq} makes the critical path 14--20\% shorter. For this design the gain in area is limited by the design itself: an 8-bit Booth multiplier that is already tight, and a chip where arithmetic is less than half of the area. The measurement setup is open and can be reused to compare any arithmetic block on the real Tiny Tapeout flow.
 
%\bibliographystyle{unsrt}
%\bibliography{refs}

\end{document}